\RequirePackage{ifpdf}
\RequirePackage{color}
\documentclass{JHEP3}

\usepackage{amsmath}
\usepackage{amssymb}
\usepackage{cancel}
\usepackage{slashed}
\usepackage{graphicx} 

\newcommand{\bea}{\begin{eqnarray}}
\newcommand{\eea}{\end{eqnarray}}
\newcommand{\bi} {\begin{itemize}}
\newcommand{\ei} {\end{itemize}}

\def\nl{\nonumber \\}

\def\d{\partial}
\def\n{\nabla}

\def\l{\left(}
\def\r{\right)}
\def\gh{{\widehat g}}
\def\gf{{ g}}

\title{
The dilaton Wess-Zumino action in higher dimensions
}

\author{
Florent Baume${}^1$ and Boaz Keren-Zur${}^1$
\\
${}^1$Institut de Th\'eorie des Ph\'enom\`enes Physiques, EPFL,\\
CH-1015 Lausanne, Switzerland\\
E-mails:  \email{florent.baume@epfl.ch}, \email{boaz.kerenzur@epfl.ch}
}

\abstract{
We present a general formula for the Wess-Zumino action associated with the Weyl anomaly, given in a curved background for any even number of dimensions.
The result is obtained by considering a finite Weyl transformation of counterterms in dimensional regularization. 
}


\begin{document}

 \section{Introduction}
\label{sec_intro}

The dilaton is a powerful probe of quantum field theories. It was used to obtain non-trivial constraints on the renormalization of unitary four dimensional theories, such as the weak irreversibility of the RG flow (the $a$ theorem\cite{Komargodski:2011vj,Cardy:1988cwa}), and the vanishing of the $\beta$ function in the asymptotic limits of perturbative RG flows\cite{Luty:2012ww}.

An important ingredient in the analyses leading to these results is the effective action of the dilaton in four dimensional CFTs. 
As part of a program aiming to generalize this methodology to higher dimensions, 
we present in this work a new formula for the dilaton Wess-Zumino action. This formula is valid for CFTs in a curved background in an arbitrary even number of dimensions.

But what is the dilaton? What is the physical meaning of its effective action?
In this work we follow the approach of \cite{Luty:2012ww} where the dilaton field, denoted by $\tau$, is defined simply as a component of a background metric $\widehat{g}^{\mu\nu}$, which is introduced via the redundant notation
\bea
	\widehat{g}^{\mu\nu}=e^{2\tau}g^{\mu\nu}~.
\eea
This notation has two advantages --- first, it allows us to conveniently realize the Weyl transformation of the metric as shifts of $\tau$
 \bea
 \label{eq_Weyl_transformation}
		\gh^{\mu\nu}\longrightarrow e^{-2\sigma}\gh^{\mu\nu}
\qquad \Rightarrow \qquad \tau \longrightarrow \tau -\sigma~, \qquad \gf^{\mu\nu}\longrightarrow \gf^{\mu\nu}
\eea
and second, in this picture the dilaton can be understood as a background field which acts as a source for the trace of the stress-energy tensor $T$
\bea
\frac{\delta }{\delta\tau(x)} W[\gh] =
2\gh^{\mu\nu} 
\frac{\delta }{\delta \gh^{\mu\nu}(x)}W[\gh]=\sqrt {-\gh}\langle  T(x)\rangle
\eea
where $W[\gh]$ is the quantum effective action\footnote{
This action can be defined via $e^{iW[\gh]}=\int D[\psi] e^{iS[\gh,\psi]}$ where $\psi$ are the dynamical fields in the theory.} written in terms of the metric $\gh$.
This definition of the dilaton makes it clear that it is not a new independent degree of freedom, and that the quantum effective action for the dilaton is nothing but a convenient bookkeeping device for correlators of $T$.

 An important observation is that this effective action is determined at a conformal fixed point 
 by the structure of the Weyl anomaly. In 4 dimensions the anomaly is given by\cite{Deser:1976yx}
 \begin{equation}\label{WeylAnomaly4D}
- \frac{1}{\sqrt {-\gh}} \delta_\sigma W[\gh] 
= \sigma \left<T(x)\right>=\sigma (c\widehat{\mathcal{W}}^2 - a\widehat{ E}_{4})
\end{equation}
where $\widehat E_4$ is the Euler density in 4 dimensions and $\widehat{\mathcal{W}}^2$ is the Weyl tensor squared, both given in terms of the metric $\gh^{\mu\nu}$. In principle the anomaly \eqref{WeylAnomaly4D} could also contain a term proportional to $\widehat{\nabla}^2\widehat{R}$. This term can, however, be understood as the variation a local $\widehat{R}^2$ term under the transformation \eqref{eq_Weyl_transformation}, and is thus not a genuine anomaly. Contributions of this type ---and their generalization to higher dimensions --- to the effective action can be eliminated by an appropriate choice of scheme and will not be considered here.

In order to find the dilaton effective action, we separate the effective action $W[\gh]$  into a part which depends on the Weyl invariant metric $\gf^{\mu\nu}$ and a part which depends also on the dilaton
\bea	
\label{eq_WZ_def}
W[\widehat{g}]={W}[\gf]+\Gamma[\gf,\tau]
\eea
Since $\gf^{\mu\nu}$ is invariant under the Weyl symmetry, we find that the $\Gamma[g_{\mu\nu},\tau]$ must be a functional
whose variation under shifts of $\tau$ gives the anomaly. It is a unique functional known as the Wess-Zumino (WZ) term. In 4 dimensions this was found to be \cite{Tomboulis:1988gw,Schwimmer:2010za}. In 4 dimensions this was found to be \cite{Fradkin:1983tg,Riegert:1984kt,Tomboulis:1988gw,Schwimmer:2010za}
\bea
\Gamma[g,\tau]&=&\int \sqrt {-\gf}d^4x \l c\tau \mathcal{W}^2 - a \l \tau E_4   +4G^{\mu\nu}\d_\mu\tau\d_\nu\tau -4 \n^2 \tau (\d\tau)^2 + 2(\d\tau)^4\r \r 
\eea
where $G^{\mu\nu}$ is the Einstein tensor, and the gravitational terms in this formula are functions of the metric $\gf^{\mu\nu}$.
Allowing for scheme dependent terms in the anomaly, such as a $\widehat{\nabla}^2\widehat{R}$ in four dimensions, one finds more interactions in the dilaton effective action. In the proof of the $a$ theorem in four dimensions such terms are eliminated from the dilaton scattering amplitude by using the on-shell condition, but in higher dimensions this is not possible, a fact that complicates the attempts to generalize the proof to higher dimensions \cite{Elvang:2012st,Elvang:2012yc}. 
 We choose our scheme such that these terms do not appear in the effective action in order to isolate the Wess-Zumino action.

Notice that under the Weyl symmetry \eqref{eq_Weyl_transformation} only the dilaton transforms. Alternatively, we could define a different symmetry:
\bea
 \label{eq_other_weyl}
		\gf^{\mu\nu}\longrightarrow e^{-2\alpha}\gf^{\mu\nu}~,
\qquad \tau \longrightarrow \tau +\alpha \qquad \Rightarrow \qquad \gh^{\mu\nu}\longrightarrow \gh^{\mu\nu}
\eea
In this case eq. \eqref{eq_WZ_def} leads to
\bea
\label{eq_other_WZ_definition}
\delta_\alpha \Gamma [\gf, \tau]=-\delta_\alpha W[\gf]=\alpha\l c \mathcal{W}^2 - a{E}_{4} \r ~.
\eea
This means that one can define the WZ action as the term whose variation under \eqref{eq_other_weyl}, in which the metric transforms as well as the dilaton, gives the anomaly written in terms of $\gf$. This is the definition which is commonly used in the literature (e.g. \cite{Komargodski:2011vj} and \cite{Elvang:2012st}),
but it is less suited for the procedure which we will describe below.

How can one extract information about the RG flow from this action? Using the effective interactions appearing in the WZ action (and imposing a clever kinematic regime known as the "on-shell" condition), it was possible to relate the "forward scattering amplitude" of the dilaton to the anomaly coefficient $a$. Writing a dispersion relation for this amplitude, and using the unitarity of the theory, the following constraint was found\cite{Komargodski:2011vj}
\bea
a_{UV}- a_{IR} > 0
\eea
where $a_{UV}$ (resp. $a_{IR}$) is the anomaly coefficients in the CFTs describing the UV (resp. IR) fixed points. 
This inequality is known as the $a$ theorem, and it implies that the RG flow between two conformal fixed points is one directional. 

One generalization of this method it to go off-criticality.
A systematic approach for computing the dilaton effective action in non-conformal theories is presented in \cite{Baume:2013}.
Such an effective action was used in \cite{Luty:2012ww} and \cite{Fortin:2012hn} to study perturbative RG flows, and to constrain their asymptotic behavior.

Another possible direction is to consider CFTs in higher dimensions, and check whether a similar analysis can be made there as well. 
Such attempts were discussed in \cite{Elvang:2012st} and \cite{Elvang:2012yc}
 for 6 and 8 dimensions respectively\footnote{An earlier derivation of the WZ action in 6 dimensions was given in \cite{Arakelian:1995ye}.}.
These attempts did not lead to a proof for an analogue of the $a$ theorem in higher dimensions, mainly due to difficulties in 
constructing dispersion relations with positivity constraints in which the contribution of the anomaly dependent terms can be isolated.

As a part of this line of research, we present here a new formula for the Wess-Zumino action in arbitrary even dimensions.
It is based on the classification of Weyl anomalies given in \cite{Deser:1993yx}, 
where it has been shown that the most general Weyl anomaly in an even number of dimensions $D=2p$ is
\begin{equation}\label{WeylAnomaly}
	\left<T(x)\right>=\sum_{i=1} c_i\widehat{I}_i - a(-1)^p\widehat{E}_{2p} 
\end{equation}
where $\sqrt {-\gh}\widehat{I}_i$ is a set of Weyl invariant scalars, $\widehat{E}_{2p}$ the Euler density and the coefficients $c_i$ and $a$ are model dependent numbers. In the terminology of \cite{Deser:1993yx} the $I_i$ anomalies, which are associated with $\ln \frac {\mu^2}{\Box}$ terms appearing in $W[\gh]$, are referred to as type B, while the Euler density anomaly is referred to as type A. In general there could be additional terms in the variation which can be eliminated by an appropriate choice of scheme. These are not genuine anomalies and will not be discussed here.

The final result of our computation is 
\bea
\label{WZfinal}
	\Gamma[g,\tau]
	&=&\int d^{2p}x\sqrt {-g} \tau \l \sum_{i=1} c_i{I}_i-a(-1)^p{E}_{2p}\r \nl
	&&-a\int d^{2p}x\sqrt {-g} \sum_{n=0}^{p-1}\sum_{k=0}^{p-n}C(p,n,k)\delta^{\mu_1\nu_1\dots\mu_n\nu_n\rho_1\dots\rho_k}_{a_1b_1\dots a_nb_nc_1\dots c_k}\
	(\partial\tau)^{2(p-n-k)}
	\prod_{i=1}^nR_{\mu_i\nu_i}^{a_ib_i}
	\prod_{j=1}^{k}\nabla_{\rho_j}\d^{c_j}\tau \nl
\eea
where
\begin{equation*}
	C(p,n,k)=
		\begin{cases}
			0&\text{if $n=p-1$, $k=1$}\\
		 (-2)^{k-n}\dfrac{(2p-2n-k-2)!p!}{n!(p-n-k)!k!}&\text{else}
		\end{cases}
\end{equation*}
The gravitational terms here are given in terms of the metric $g^{\mu\nu}$, we used $\n$ to denote  the covariant derivative, and the generalized Kronecker delta was defined via
\begin{equation}
	\delta^{\mu_1\ldots\mu_n}_{a_1\ldots a_n}=n!e^{\phantom{[}\mu_1}_{[a_1}\ldots e^{\mu_n}_{a_n]}~.
\end{equation}

The computation leading to this result is described in section \ref{sec_computation}, and some useful definitions and formulae are given in appendices \ref{app_Weyl} and \ref{app_Kronecker}. Explicit expressions for the action in 2,4 and 6 dimensions are given in appendix \ref{app_results}.

 \section{The computation}
\label{sec_computation}

 \subsection{The main idea}

In this section we will show how to compute the Wess-Zumino (WZ) action in dimensional regularization ($D=2p+\epsilon$).
Our computation is similar in spirit to the one presented in \cite{Tomboulis:1988gw} (a similar computation was presented recently in \cite{Coriano:2013xua}), but here we use this approach to find the 
WZ action for arbitrary, even, number of dimensions.

The main point is that in dimensional regularization the anomaly comes about from the non-invariance of counterterms.
A theory which is classically Weyl invariant can be regularized in a symmetry preserving way, 
but in order to have a finite theory in the $\varepsilon\to0$ limit we must add local counterterms, and these counterterms break the symmetry explicitly.
The finite part in the variation of these counterterms is the anomaly.
Generalizing eq. (2.3) of \cite{Duff:1977ay} to higher dimensions, we find that the form of the counterterms
which leads to the anomaly \eqref{WeylAnomaly} is
\bea
\label{eq_anomaly_counterterms}
W[\gh]&\supset&W_{CT}[\gh]=\int d^Dx \sqrt {-\gh} \frac{\mu^{-\varepsilon}}{\epsilon}
\l -\sum_{i=1} c_i\widehat{I}_{i,D}+a(-1)^p\widehat{E}_{2p}\r 
\eea
where $\widehat{I}_{i,D}$ are a set of Weyl invariant scalars, defined in the $D$ dimensional theory and which coincide with $\widehat{I}_{i}$ when $D=2p$, and $\widehat{E}_{2p}$ is the Euler density defined in $2p$ dimensions. 
Indeed, these terms lead to a finite anomaly because the Weyl variations of the scalars appearing there contain evanescent terms
\bea
\delta_\sigma \sqrt {-\gh} \widehat{I}_{i,D} &=& \sigma\varepsilon \sqrt {-\gh} \widehat{I}_{i,D}\nl
\delta_\sigma \sqrt {-\gh}\widehat  E_{2p} &=&  \sigma\varepsilon \sqrt {-\gh}  \widehat E_{2p} ~.
\eea
It is important for our discussion that the remaining terms in the effective action are Weyl invariant, so we define
\bea
W_{inv}[\gh]=W[\gh]-W_{CT}[\gh]~.
\eea

Let us now use the above discussion to compute the dilaton effective action defined via 
\bea
\label{eq_WZ_definition}
	W[\widehat{g}]={W}[g]+\Gamma[ g,\tau]~.
\eea
The Weyl variation of the LHS gives the anomaly, and it is obvious that in the RHS only $\Gamma[ g,\tau]$ can be anomalous (recall that we defined $\gf^{\mu\nu}$ to be invariant under the Weyl transformation). It is also clear that $W_{inv}[\gh]=W_{inv}[\gf]$. Subtracting these terms from eq. \eqref{eq_WZ_definition} we find
\bea
\label{eq_important}
\Gamma[g,\tau]&=&\l W_{CT}[\widehat{g}]-W_{CT}[g]\r _{\varepsilon=0}~.
\eea
This is the basic formula from which we derive the result, and from here on all that is left are the technical aspects of the computation.

 \subsection{Results}
\subsubsection{The type B anomalies}
A simple application of eq. \eqref{eq_important} is the case of the type B anomalies.
Using the relation
\bea
\sqrt {-\gh} \widehat{I}_{i,D}&=&e^{-\varepsilon \tau}\sqrt {-\gf} {I}_{i,D}
\eea
we find that the WZ action for the type B anomaly (denoted by $\Gamma_B$) is 
\bea
\Gamma_B [g,\tau]&=&\left[
\int d^Dx \sqrt {-\gf} \frac{\mu^{-\varepsilon}}{\varepsilon}(-e^{-\varepsilon \tau}+1) \sum_{i=1} c_i{I}_{i,D} \right] _{\varepsilon=0}\nl
&=&\int d^{2p}x \sqrt {-\gf} \tau \sum_{i=1} c_i{I}_{i}
\eea

\subsubsection{The type A anomaly}
The case of the type $A$ anomaly, proportional to the Euler density, is more complicated. As in the case of the type $B$ anomaly, we would like to write $\sqrt {-\gh} \widehat E_{2p}$  in terms of $\gf$ and $\tau$
\bea
\sqrt {-\gh} \widehat E_{2p}
&=&e^{-\varepsilon \tau}\sqrt {-\gf}\l E_{2p} + \delta_\tau E_{2p}\r 
\eea
The function $\delta_\tau E_{2p}$ can be expanded in powers of $\varepsilon$
\bea
\sqrt {-\gh} \widehat E_{2p}
&=&e^{-\varepsilon \tau}\sqrt {-\gf}\l E_{2p} + (\delta_\tau E_{2p})\big |_{\varepsilon=0}+\varepsilon \frac {\d}{\d \varepsilon} (\delta_\tau E_{2p}) \big |_{\varepsilon=0}
+O(\epsilon^2)\r ~.
\eea
Notice that in the $\varepsilon=0$ limit $E_{2p}$ is a total derivative, therefore its Weyl variation must be a total derivative as well. Denoting $(\delta_\tau E_{2p})\big |_{\varepsilon=0} = \d_\mu V^\mu$ and plugging into eq. \eqref{eq_important} we find
\bea
\label{eq_WZ_structure}
\Gamma_A[g,\tau]&=&
a(-1)^{p}\left [\int d^Dx \sqrt {-\gf} \frac{\mu^{-\varepsilon}}{\varepsilon}\l\l e ^{-\varepsilon \tau}-1 \r  E_{2p} +e^{-\varepsilon \tau} \d_\mu V^\mu
+e^{-\varepsilon \tau}\varepsilon \frac {\d}{\d \varepsilon} (\delta_\tau E_{2p}) \big |_{\varepsilon=0}\r   \right] _{\varepsilon=0}\nl
&=&
 -a(-1)^{p}\int d^{2p}x \sqrt {-\gf}  \l 
\tau E_{2p} - \d_\mu \tau V^\mu   -\frac {\d}{\d \varepsilon} (\delta_\tau E_{2p})\r +(total~derivatives)\nl
\eea

Finding the functions $V^\mu$ and $\frac {\d}{\d \varepsilon} (\delta_\tau E_{2p})$ 
is a nontrivial task, and here we will only sketch the outline of the computation.
We begin with the definition of the Euler density: 
\begin{equation}
	 \widehat{E}_{2p}=\frac{1}{2^p}\widehat\delta^{\mu_1\nu_1\dots\mu_p\nu_p}_{a_1b_1\dots a_pb_p}
	\prod_{i=1}^p\widehat{R}_{\mu_i\nu_i}^{~~~~a_ib_i}
\end{equation}
where $\widehat \delta^{\mu_1\nu_1\dots\mu_p\nu_p}_{a_1b_1\dots a_pb_p}$ is the generalized Kronecker delta defined as the antisymmetrization of the vielbeins $\widehat {e}\;_\mu^a$ (see equation \eqref{def:generalizedKronecker} in the appendix).

Using the expression of $\widehat{R}_{\mu\nu}^{\phantom{\mu\nu}ab}$ in terms of the curvature of the metric $g^{\mu\nu}$ and the dilaton
\begin{equation}
	\widehat{R}_{\mu \nu}^{\phantom{\mu\nu}ab}=R_{\mu \nu}^{\phantom{\mu\nu}ab}+4e^{[a}_{[\mu}\nabla^{\vphantom{[}}_{\nu]}\partial^{b]}\tau+4e^{[a}_{[\mu}\partial^{\vphantom{]}}_{\nu]}\tau\partial^{b]}\tau-2e^a_{[\mu}e^b_{\nu]}(\partial\tau)^2~,
\end{equation}
and eq. \eqref{contrDelta} we find
\begin{align}
	\sqrt{-\widehat{g}}\widehat{E}_{2p}
	=&e^{-\varepsilon\tau}\sqrt{-g}\sum_{n=0}^p\sum_{k=0}^{p-n}
	\Gamma(D-2n-k+1)\frac{(-1)^{p-n-k}2^{k-n}p!}{n!(p-n-k)!k!}\delta^{\mu_1\nu_1\dots\mu_n\nu_n\rho_1\dots\rho_k}_{a_1b_1\dots a_nb_nc_1\dots c_k}\nl
		&\times(\partial\tau)^{2(p-n-k)}
		\prod_{i=1}^nR_{\mu_i\nu_i}^{a_ib_i}
		\prod_{j=1}^{k}\bigg(
			\nabla^{\vphantom{[}}_{\rho_j}\partial^{c_j}\tau+\partial^{\vphantom{]}}_{\rho_j}\tau\partial^{c_j}\tau
		\bigg)~.
\end{align}

The next step is to notice that because of antisymmetry, we cannot have expressions proportional to $\partial_{\rho_i}\tau\partial^{c_i}\tau\partial_{\rho_j}\tau\partial^{c_j}\tau$. Removing such terms and using equation \eqref{eq_useful}, we can identify a term linear in $\varepsilon$, from which we extract
\bea
\label{eq_V}
	\frac {\d}{\d \varepsilon} (\delta_\tau E_{2p}) \big|_{\varepsilon=0}
	&=&\sum_{n=0}^{p-1}\sum_{k=0}^{p-n} 2^{k-n}(-1)^{p-n-k} \frac{(2p-2n-k-1)!p!}{n!(p-n-k)!k!}\delta^{\mu_1\nu_1\dots\mu_n\nu_n\rho_1\dots\rho_k}_{a_1b_1\dots a_nb_nc_1\dots c_k}\nl
&&\times	(\partial\tau)^{2(p-n-k)}
	\prod_{i=1}^nR_{\mu_i\nu_i}^{a_ib_i}
	\prod_{j=1}^{k}\nabla_{\rho_j}\d^{c_j}\tau
\eea
In order to identify the $\varepsilon$ independent terms as total derivatives we use the the Bianchi identity and the definition of the Riemann tensor
\bea
\label{def:RiemannTensor}
\nabla_{[\rho}^{\phantom{a}} R_{\mu\nu]}^{\phantom{\mu\nu}ab}&=&0\nl
2\nabla_{[\mu}\nabla_{\nu]}\partial^a\tau&=&R_{\mu\nu}^{\phantom{\mu\nu}ab}\partial_b^{\vphantom{a}}\tau~.
\eea
After some algebra we find
\bea	
\label{eq_D}
	 (\delta_\tau E_{2p})\big |_{\varepsilon=0}
	 &=&\sum_{n=0}^{p-1}\sum_{k=1}^{p-n}2^{k-n}(-1)^{p-n-k} \frac{(2p-2n-k-1)! p!}{n!(p-n-k)!(k-1)!}\delta^{\mu_1\nu_1\dots\mu_n\nu_n\rho_1\dots\rho_k}_{a_1b_1\dots a_nb_nc_1\dots c_k}\nl
	&&\times\nabla_{\rho_k}\bigg(\partial^{c_k}\tau (\partial\tau)^{2(p-n-k)}\prod_{i=1}^nR_{\mu_i\nu_i}^{a_ib_i}\prod_{j=1}^{k-1}\nabla_{\rho_j}\d^{c_j}\tau \bigg)~,
\eea
which, as expected, is a total derivative.
Combining equations \eqref{eq_V} and \eqref{eq_D} into \eqref{eq_WZ_structure},  we find the formula quoted in \eqref{WZfinal}.
This formula agrees with the known results for up to 8 dimensions\cite{Elvang:2012st, Elvang:2012yc}.

\section{Conclusions}
We have used dimensional regularization to find a general formula for the WZ action in any even number of dimensions. 
Let us make some observations regarding this result. 
\begin{itemize}
\item
An interesting property of this result is that the $k=0$ terms are proportional to the Euler densities in lower dimensions
\bea
	\Gamma[g,\tau]
	&\supset&-a \int d^4x\sqrt {-g}
	\sum_{n=0}^{p-1}
	(-1)^{n}2^{p-n}\dfrac{(2p-2n-2)!p!}{n!(p-n)!}
	(\partial\tau)^{2(p-n)}E_{2n}
\eea
 \item
In $D>2$ it is useful to write the action in terms of the conformal compensator $\Omega=e^{-(p-1)\tau}$.
The main advantage of this presentation is that it makes it easier to impose the "on-shell" condition $\Box\Omega=0$,
a kinematic constraint that in four dimensions was shown to clean the effective action from improvement dependent effects\cite{Luty:2012ww}.
We find
\bea
	\Gamma[g,\Omega]
	&=&-\frac{1}{p-1}\int d^{2p}x\sqrt {-g} \ln \Omega \l \sum_{i=1} c_i{I}_i-a(-1)^p{E}_{2p}\r \nl
	&&-\,a \int d^{2p}x\sqrt {-g}\sum_{n=0}^{p-1}\sum_{k=0}^{p-n}
\frac{(-1)^k }{(p-1)^{(2p-2n-k)}}C(p,n,k)
\delta^{\mu_1\nu_1\dots\mu_n\nu_n\rho_1\dots\rho_k}_{a_1b_1\dots a_nb_nc_1\dots c_k}\Omega^{-(2p-2n-k)} (\partial\Omega)^{2(p-n-k)}\nl
	&&
		~~~~~~~~~~~~~~~~~~~~~~~~~~~\times \prod_{i=1}^nR_{\mu_i\nu_i}^{a_ib_i}
  		\bigg(
			\nabla_{\rho_1}\d^{c_1}\Omega-k\Omega^{-1}\partial_{\rho_1}\Omega\partial^{c_1}\Omega
		\bigg)\prod_{j=2}^{k}\nabla_{\rho_j}\d^{c_j}\Omega
\eea
where we used
\bea
	\partial^a \tau&=&-\frac {1}{p-1}\Omega^{-1}\partial^a\Omega\nl
	\nabla_\mu\d^a\tau&=&\frac {1}{p-1}\l \Omega^{-2}\partial_\mu\Omega\partial^a\Omega-\Omega^{-1}\nabla_\mu\d^a\Omega\r
\eea
\item
It is shown in \cite{Elvang:2012yc}, using the terminology of GJMS operators, that the WZ action in flat space can be rewritten, after imposing the on-shell condition, as 
\bea
\Gamma_{A}[\gf=\eta,\tau]&\propto&\tau \Box^{D/2}\tau~.
\eea 
In principle, it should be possible to derive this result from our general equation \eqref{WZfinal} by multiple integration by parts, and substitution of the on-shell condition.
This is a relatively technically involved computation which we leave for a separate project.

  It is shown in \cite{Elvang:2012yc}, using the terminology of GJMS operators, that the WZ action in flat space can be rewritten --- after adding invariant terms under the transformation \eqref{eq_other_weyl} and imposing the on-shell condition --- as 
\bea
\Gamma_{A}[\gf=\eta,\tau]&\propto&\tau \Box^{D/2}\tau~.
\eea 
In principle, it should be possible to derive this result from our general equation \eqref{WZfinal} by adding local terms to the effective action (such as $\widehat{R}^2$ for $D=4$), multiple integration by parts, and substitution of the on-shell condition.
It would also be interesting to find a general way of getting the GJMS operators using our result.
Those are relatively technically involved computations which we leave for a separate project.

  It is shown in \cite{Elvang:2012yc}, using the terminology of GJMS operators, that the WZ action in flat space can be rewritten --- after adding invariant terms under the transformation \eqref{eq_other_weyl} and imposing the on-shell condition --- as 
\bea
\Gamma_{A}[\gf=\eta,\tau]&\propto&\tau \Box^{D/2}\tau~.
\eea 
In principle, it should be possible to derive this result from our general equation \eqref{WZfinal} by adding local terms to the effective action (such as $\widehat{R}^2$ for $D=4$), multiple integration by parts, and substitution of the on-shell condition.
This is a relatively technically involved computation which we leave for a separate project.

\end{itemize}

 \section*{Acknowledgments}

We thank R. Rattazzi for useful discussions.
The research was supported by the Swiss National Science Foundation under grants 200020-138131 and 200021-125237.

\appendix

\section{Geometry and Weyl transformations}
\label{app_Weyl}
Under Weyl transformation, the metric and vielbeins change as
\begin{equation}\label{Weylappendix}
	\left\{
		\begin{aligned}
			\widehat{g}^{\mu\nu}&\longrightarrow {\widehat{g}\,'}^{\mu\nu}= e^{-2\sigma}\widehat{g}^{\mu \nu}\\
			\widehat{e}\;_a^\mu&\longrightarrow {\widehat{e}\,'}_a^{\mu}=e^{-\sigma} \widehat{e}\;_a^\mu
		\end{aligned}
	\right.
\end{equation}
With this definition one can see the measure of the integration changes as:
\begin{equation}\label{sqrtgWeyl}
	\sqrt{-\widehat{g}}\longrightarrow \sqrt{-\widehat{g}\;'} =e^{\sigma D}\sqrt{-\widehat{g}}
\end{equation}

Our convention for the Riemann and Ricci curvatures is 
\bea
[\widehat {\nabla}_\mu,\widehat \nabla_\nu]V_\rho= \widehat R_{\mu\nu\rho}^{\phantom{\mu\nu\rho}\sigma}V_\sigma
\qquad\qquad
  \widehat R_{\mu\nu}= \widehat R_{\mu\alpha\nu}^{\phantom{\mu\nu\alpha}\alpha}~.
\eea
where the Riemann tensor and covariant derivative $\widehat \n$ are given in terms of the metric $\gh$.
The Riemann tensor transforms as follows:
\begin{equation}\label{Rw}
	\widehat{R}_{\mu \nu}^{\phantom{\mu\nu}ab}\longrightarrow{\widehat{R}\,'}_{\mu \nu}^{\phantom{\mu\nu}ab}=\widehat{R}_{\mu \nu}^{\phantom{\mu\nu}ab} -4e^{[a}_{[\mu}\widehat{\nabla}^{\vphantom{[}}_{\nu]}\partial^{b]}\sigma+4e^{[a}_{[\mu}\partial^{\vphantom{]}}_{\nu]}\sigma\partial^{b]}\sigma-2e^a_{[\mu}e^b_{\nu]}(\partial\sigma)^2
\end{equation}
Equation \eqref{Rw} can also be adapated to extract the dilaton from the Riemann tensor:
\begin{align}
	\widehat{R}_{\mu \nu}^{\phantom{\mu\nu}ab}=R_{\mu \nu}^{\phantom{\mu\nu}ab}+4e^{[a}_{[\mu}\nabla^{\vphantom{[}}_{\nu]}\partial^{b]}\tau+4e^{[a}_{[\mu}\partial^{\vphantom{]}}_{\nu]}\tau\partial^{b]}\tau-2e^a_{[\mu}e^b_{\nu]}(\partial\tau)^2\label{Riemhat}\\
	\widehat{R}=e^{2\tau}\bigg(R+2(D-1)\nabla^\alpha\partial_\alpha\tau-(D-1)(D-2)(\partial\tau)^2\bigg)\label{Rhat}
\end{align}
where $R_{\mu \nu}^{\phantom{\mu\nu}ab}$ and $\n$ are the Riemann tensor and the covariant derivative associated with the metric $\gf$.


\section{The generalized Kronecker delta}
\label{app_Kronecker}

To work with the Euler anomaly in arbitrary even dimension, we will need to use the generalized Kronecker delta defined as:
\begin{equation}\label{def:generalizedKronecker}
	\delta^{\mu_1\mu_2\ldots\mu_n}_{a_1a_2\ldots a_n}=n!e^{\phantom{[}\mu_1}_{[a_1}\ldots e^{\mu_n}_{a_n]}
\end{equation}
This object satisfies the following useful relations:
\begin{itemize}

	\item The contraction of the generalized Kronecker delta with a vielbein is given by
		\begin{equation*}
			\delta^{\mu_1\ldots\mu_n}_{a_1\ldots a_n}e^{a_n}_{\mu_n}=(D-n+1)\delta^{\mu_1\ldots\mu_{n-1}}_{a_1\ldots a_{n-1}}~,
		\end{equation*}
		which is easily generalized to:
		\begin{equation}
			\delta^{\mu_1\ldots\mu_m\ldots\mu_{m+n}}_{a_1\ldots a_m\ldots a_{m+n}}\prod_{i=1}^ne^{a_i}_{\mu_i}=
			\frac{(D-m)!}{(D-m-n)!}\delta^{\mu_1\ldots\mu_m}_{a_1\ldots a_m}~.
		\end{equation}
		If $m+n=2p$ and $D-2p< 1$ this can be written as
		\begin{equation}\label{contrDelta}
			\delta^{\mu_1\ldots\mu_m\ldots\mu_{m+n}}_{a_1\ldots a_m\ldots a_{m+n}}\prod_{i=1}^{n}e^{a_i}_{\mu_i}=\Gamma(D-m+1)\delta^{\mu_1\ldots\mu_m}_{a_1\ldots a_m}~.
		\end{equation}
		
	\item Under Weyl transformation:
	\bea
	\delta_\sigma \sqrt {-\gh} \widehat \delta^{\mu_1\ldots\mu_{n}}_{a_1\ldots a_{n}}&=&e^{(D-n)\sigma}\sqrt {-\gh} \widehat \delta^{\mu_1\ldots\mu_{n}}_{a_1\ldots a_{n}}
	\eea

	\item Another useful property is the contraction of two indices with a tensor $A^a_\mu$, while all the other indices are contracted with a product of tensors $B^b_\nu$. First, one can find
		\begin{align*}
			\delta^{\mu_1\dots\mu_p\nu}_{a_1\dots a_pb}=e^\nu_b\delta^{\mu_1\dots\mu_p}_{a_1\dots a_p}-\sum_{k=1}^p e^\nu_{a_k}\delta^{\mu_1\dots\mu_k\dots\mu_p}_{a_1\dots b\dots a_p}
		\end{align*}
		It immediatly follows that
	\begin{align}\label{ProdDelta}
		\delta_{b_1\ldots b_q a}^{\nu_1\ldots \nu_q \mu}A^a_{\mu}\prod_{i=1}^q B_{\nu_i}^{b_i}
		=
		\delta^{\nu_1\ldots\nu_q}_{b_1\ldots b_q}\prod_{i=1}^{q-1} B_{\nu_i}^{b_i}\left(
			B_{\nu_q}^{b_q}A^\mu_{\mu}-q A^{b_q}_{\mu}B_{\nu_q}^{\mu}
		\right)~.
	\end{align}
	
		To generalize this formula to more tensors, notice that 
		\begin{align*}
			\delta^{\mu_1\dots\mu_p\nu_1\dots\nu_q\rho_1\dots\rho_r\dots\sigma}_{a_1\dots a_pb_1\dots b_q c_1\dots c_r\dots d}
			=&e^\sigma_d\delta^{\mu_1\dots\mu_p\nu_1\dots\nu_q\rho_1\dots\rho_r\dots}_{a_1\dots a_pb_1\dots b_q c_1\dots c_r\dots}
			-\sum_{k=1}^pe^\sigma_{a_k}\delta^{\mu_1\dots\mu_k\dots\mu_p\nu_1\dots\nu_q\rho_1\dots\rho_r\dots}_{a_1\dots d\dots a_pb_1\dots b_q c_1\dots c_r\dots}\\
			&-\sum_{k=1}^qe^\sigma_{b_k}\delta^{\mu_1\dots\mu_p\nu_1\dots\nu_k\dots\nu_q\rho_1\dots\rho_r\dots}_{a_1\dots a_pb_1\dots d \dots b_q c_1\dots c_r\dots}-\dots
		\end{align*}
	Thus 
\bea	
\label{eq_useful}
	\delta^{\mu_1\nu_1\dots\mu_p\nu_p\rho_1\dots\rho_q\sigma}_{a_1b_1\dots a_pb_pc_1\dots c_q d}
		\prod_{i=1}^pA_{\mu_i\nu_i}^{a_ib_i}\prod_{j=1}^qB_{\rho_j}^{c_j}C^\sigma_d
		&=&\delta^{\mu_1\nu_1\dots\mu_p\nu_p\rho_1\dots\rho_q}_{a_1b_1\dots a_pb_pc_1\dots c_q}
		\prod_{i=1}^{p-1}A_{\mu_i\nu_i}^{a_ib_i}\prod_{j=1}^{q-1}B_{\rho_j}^{c_j}\label{Laplace3}\\
		&&\times\bigg(
			A_{\mu_p\nu_p}^{a_pb_p}B_{\rho_q}^{c_q}C_\alpha^\alpha
			-2pA_{\mu_p\nu_p}^{\alpha b_p}B^{c_q}_{\rho_q}C_\alpha^{a_p}
			-qA_{\mu_p\nu_p}^{a_pb_p}B_{\rho_q}^\alpha C_\alpha^{c_q}
			\bigg)\nonumber
	\eea
\end{itemize}

\section{Explicit results in $D=2,4,6$}
\label{app_results}
We give here the expression of the type A part of the Wess-Zumino term up to $p=3$.
We give the results in two forms -- the one given by applying eq. \eqref{WZfinal} directly, and another one which is obtained after integration by parts.
\begin{itemize}
\item $D=2$
\bea
\Gamma_{A,2}[g,\tau]&=&a\int d^2x\sqrt {-\gf} \l  \tau R   - (\d\tau)^2 \r 
\eea
(in $D=2$ the coefficient of the type A anomaly is denoted by $c$).
\item $D=4$
\bea
  \Gamma_{A,4}&=&-a\int d^4x\sqrt{-g}\bigg(\tau E_4-4\nabla_\mu\d_\nu\tau\nabla^\mu\d^\nu\tau+4(\nabla^2\tau)^2-2R(\partial\tau)^2
  	-4\nabla^2\tau(\partial\tau)^2+2(\partial\tau)^2\bigg)\nl
	&=&-a\int d^4x\sqrt{-g}\bigg(\tau E_4+4G_{\mu\nu}\partial^\mu\tau\partial^\nu\tau-4\nabla^2\tau(\partial\tau)^2+2(\partial\tau)^2
		\bigg)
\eea

\item $D=6$
\bea
  \Gamma_{A,6}=&a{\displaystyle\int} d^6x\sqrt{-g}\bigg(&\tau E_6-3E_4(\partial\tau)^2+24R_{\mu\nu\rho\lambda}\nabla^\mu\d^\rho\tau\nabla^\nu\d^\lambda\tau+12R\big((\nabla^2\tau)^2-\nabla^\mu\d^\nu\tau\nabla_\mu\d_\nu\tau\big)\nl
  &&-48R_{\mu\nu}\nabla^\mu\d^\nu\tau\nabla^2\tau+48R_{\mu\nu}\nabla^\mu\d^\rho\tau\nabla_\rho\d^\nu\tau+8(\nabla^2\tau)^3-24\nabla^2\tau\nabla^\mu\d^\nu\tau\nabla_\mu\d_\nu\tau\nl
  &&+16\nabla^\mu\d^\nu\tau\nabla_\nu\d^\rho\tau\nabla_\mu\d_\rho\tau+24G_{\mu\nu}\nabla^\mu\d^\nu\tau(\partial\tau)^2+6R(\partial\tau)^4\nl
  &&-24\big( (\nabla^2\tau)^2-\nabla_\mu\d_\nu\tau\nabla^\mu\d^\nu\tau\big)(\partial\tau)^2+36\nabla^2\tau(\partial\tau)^4-24(\partial\tau)^6
  \bigg)\nl
=&a{\displaystyle\int} d^6x\sqrt {-\gf} \bigg( & \tau E_6  -3E_4 (\d\tau)^2   
+12\l R_\mu^{\phantom{\mu} \rho\lambda\sigma}R_{\nu\rho\lambda\sigma}
-2R_{\mu\rho\nu\lambda}R^{\rho\lambda}
+RR_{\mu\nu}
-2R_{\mu\rho}R^\rho_\nu\r \partial^\mu\tau\partial^\nu\tau\nl
&&-16R_{\mu\nu\rho\sigma}\nabla^\mu\d^\rho\tau\partial^\nu\tau \partial^\sigma\tau 
	+8R \l- \n^2\tau (\d\tau)^2   +\partial_\mu\tau\partial_\nu\tau \nabla^\mu\d^\nu\tau\r \nl
&&+16R_{\mu\nu}\l \nabla^\mu\d^\nu\tau(\partial\tau)^2 + \partial^\mu\tau\partial^\nu\tau\nabla^2\tau - \partial^\mu\tau\nabla^\nu(\partial\tau)^2\r 
+6R(\partial\tau)^4
\nl
&&			-24\l (\n^2\tau)^2 - \n_\mu \d_\nu \tau \n^\mu\d^\nu\tau\r(\partial\tau)^2
			+36\nabla^2\tau(\partial\tau)^4-24(\partial\tau)^6\bigg)
\eea

\end{itemize}


\end{document}